\documentclass[pra,twocolumn,aps,superscriptaddress,longbibliography]{revtex4-1}

\usepackage{amsmath}
\usepackage{amssymb}
\usepackage{graphicx}
\usepackage[usenames,dvipsnames]{color}
\usepackage{braket}
\usepackage{bm}
\usepackage{times}
\usepackage{hyperref}
\usepackage{pdfpages}
\usepackage{float}
\usepackage{lipsum}
\usepackage[dvipsnames]{xcolor}
\hypersetup{
  colorlinks=true,
  citecolor=blue,
  linkcolor=blue,
  urlcolor=blue}
\usepackage{tikz}
\usetikzlibrary{decorations.markings}
\usetikzlibrary{decorations.pathmorphing,snakes}
\usetikzlibrary{arrows.meta}
\tikzset{middlearrow/.style={
        decoration={markings,
            mark= at position 0.5 with {\arrow{#1}} ,
        },
        postaction={decorate}
    }
}

\makeatletter
 \AtBeginDocument{\let\LS@rot\@undefined}
\makeatother

\begin{document}
\title{Bosonic quantum Hall states in single layer 2D optical lattices}

\author{Rukmani Bai}
\affiliation{Physical Research Laboratory,
             Ahmedabad - 380009, Gujarat,
             India}
\affiliation{Indian Institute of Technology Gandhinagar,
             Palaj, Gandhinagar - 382355, Gujarat,
             India}
\author{Soumik Bandyopadhyay}
\affiliation{Physical Research Laboratory,
             Ahmedabad - 380009, Gujarat,
             India}
\affiliation{Indian Institute of Technology Gandhinagar,
             Palaj, Gandhinagar - 382355, Gujarat,
             India}
\author{Sukla Pal}
\affiliation{Physical Research Laboratory,
             Ahmedabad - 380009, Gujarat,
             India}
\author{K. Suthar}
\affiliation{Physical Research Laboratory,
             Ahmedabad - 380009, Gujarat,
             India}
\author{D. Angom}
\affiliation{Physical Research Laboratory,
             Ahmedabad - 380009, Gujarat,
             India}

\begin{abstract}
 Quantum Hall (QH) states of two dimensional (2D) single layer optical 
lattices are examined using Bose-Hubbard model (BHM) in presence of artificial 
gauge field. We study the QH states of both the homogeneous and inhomogeneous 
systems. For the homogeneous case we use cluster Gutzwiller mean-field (CGMF) 
theory with cluster sizes ranging from $2\times 2$ to $5\times 5$. We, then, 
consider the inhomogeneous case, which is relevant to experimental realization. In this case, we use CGMF and exact diagonalization (ED). The ED studies are 
using lattice sizes ranging from $3\times 3$ to $4\times 12$. Our results show 
that the geometries of the QH states are sensitive to the magnetic 
flux $\alpha$ and cluster sizes. For homogeneous system, among various 
combinations of $1/5\leqslant \alpha\leqslant 1/2$ and filling factor 
$\nu$, only the QH state of $\alpha=1/4$ with $\nu=1/2$, $1$, $3/2$ and $2$ 
occur as ground states. For other combinations, the competing superfluid (SF) 
state is the ground state and QH state is metastable. For BHM with envelope 
potential, all the QH states observed in homogeneous system exist for box 
potentials, but none for the harmonic potential. The QH states also persist 
for very shallow Gaussian envelope potential. As a possible experimental 
signature we study the two-point correlations of the QH and SF states.
\end{abstract}



\maketitle

\section{Introduction}

The experimental realization of Bose Einstein condensates 
(BECs) of dilute atomic gases in optical lattices 
\cite{anderson_98,greiner_01,greiner_02,lewenstein_07}, and consequent 
developments \cite{oliver_06,bloch_08} have opened new frontiers to explore 
the physics of quantum many-body systems. This is due to the possibility of 
experimental control on the inter-atomic interactions, number of atoms, lattice 
geometry and choice of atomic species. In particular, bosons in optical 
lattices are near ideal realizations \cite{jaksch_98} of the Bose-Hubbard  
model (BHM) \cite{fisher_89, hubbard_63}. The recent experimental 
implementations of artificial gauge potential 
\cite{lin_09a,lin_09b,dalibard_11,lin_11,aidelsburger_11,
aidelsburger_13, miyake_13} in optical lattices have introduced an important 
parameter and made these systems excellent testing ground for QH physics 
\cite{goldman_16}. Despite enormous progress in experimental and theoretical 
understanding of QH effect \cite{ezawa_2013,wen_92,murthy_03, hansson_17}, 
a basic understanding of the fractional quantum Hall (FQH) effect 
\cite{stormer_82} is still missing. The major difficulty arises from the 
strong correlations of electrons, but which is also the origin of FQH states. 
Although, the Laughlin ansatz \cite{laughlin_83} provides exact solutions for 
some FQH systems, but it is not yet observed in experiments. The strong 
magnetic field required to obtain FQH states is the major hurdle to observe 
these many-body states. Optical lattices, in this respect, have the advantage 
as various topological states, such as FQH states, are predicted to occur 
within the range of parameters achieved in experiments 
\cite{palmer_06, sorensen_05}. 
 
In the BHM Hamiltonian the hopping and on-site interaction are the two 
competing terms. And both of these can be tuned by changing the depth of the 
lattice potential and employing Feshbach resonance \cite{inouye_98,chin_10}. 
The hopping parameter $J$, which defines the strength of the hopping term 
in the BHM Hamiltonian, acquires a phase $J \rightarrow |J|\exp(i\Phi)$ in 
the presence of an artificial gauge potential~\cite{garcia_12} through the 
Peierls substitution \cite{peierls_33,hofstadter_76} and modifies the states 
of BHM. So, for an atom in the optical lattice there is a change of 
phase $\Phi = 2\pi\alpha$ when it hops around an unit cell or plaquette, 
where $\alpha$ is the flux quanta per plaquette. In theoretical studies, 
features of Laughlin states in low particle density limit has been 
reported \cite{hafezi_07} for $\nu =1/2$ and $\alpha < \alpha_c = 0.4$. 
Here, $\nu$ is the filling factor, the number of particles per flux quanta 
and $\alpha_c$ is the critical value below which FQH states exist. 
For  $\alpha > \alpha_c$ the equilibrium ground state properties start to 
change. And,  the existence of a striped vortex lattice phase is reported in 
the neighbourhood of $\alpha = 1/2$ \cite{palmer_08}. On the other hand, 
based on the results of Monte Carlo and exact diagonalization (ED), the 
existence of bosonic FQH states is predicted \cite{umucaliler_10} in the 
vicinity of Mott plateaus for $\alpha = 2/3$. 
Similar results are reported in a recent work using the Chern-Simons 
theory \cite{kuno_17} in combination with single site Gutzwiller mean-field 
(SGMF) theory. In another recent work \cite{natu_16}, the incompressibility 
of the FQH states is employed to identify these states in computations using 
cluster Gutzwiller mean-field (CGMF) theory for $\alpha = 1/5$ at $\nu = 1/2$. 
On the other hand, using reciprocal cluster mean-field (RCMF) analysis
H$\ddot{u}$gel {\it et al.} \cite{hugel_17} predicted a competing FQH state 
as a metastable state for $\alpha = 1/4$. In this work, we report 
FQH states at distinct $\nu$s for low and high flux. For example, 
when $\alpha = 1/5$ we obtain QH states at $\nu = n/2$, where $n = 1,2,..,9$ 
and for $\alpha = 1/2$ at $\nu = 1/2$, $1$, and $3/2$. In particular, we 
discuss the QH states for $\alpha=1/5$, $1/4$ and $1/2$ in the hard-core 
boson limit. We also obtain QH states for $\alpha=1/3$ case, however, we 
have not provided the details as the general trend is similar to 
$\alpha=1/5$. 

Motivated by the recent theoretical investigations and experimental progress, 
we address a basic gap in our current understanding. And, that is the 
occurrence of QH states in optical lattices with an envelope potential. 
This key issue is addressed in this work.  For our studies we use 
SGMF \cite{rokhsar_91,sheshadri_93,Bissbort_09} and CGMF 
\cite{Buonsante_04,Yamamoto_09,Pisarski_11,McIntosh_12,luhmann_13}
theories, and ED. Our results, for the case of homogeneous optical lattices, 
agree well with the previous theoretical observations. After establishing this 
and demonstrating that getting the geometry of QH states requires larger 
cluster sizes in CGMF, we provide an answer to the question: what is the 
nature of the QH states in optical lattices with an envelope potential?


\section{Theoretical methods}
We consider bosonic atoms at zero temperature 
confined in a two-dimensional (2D) square optical lattice with an
envelope potential in presence of synthetic magnetic 
field~\cite{jaksch_03,aidelsburger_11,aidelsburger_13,miyake_13}. 
In the Landau gauge, the system is well described by the BHM
~\cite{jaksch_98,jaksch_03,sorensen_05,palmer_06,palmer_08} 
with Peierls substitution in the nearest-neighbour (NN) hopping
~\cite{peierls_33,harper_55,hofstadter_76}, and the Hamiltonian is
\begin{eqnarray}
 \hat{H} & =& -\sum_{p, q}\left[\left (J_x {\rm e}^{i2\pi\alpha q} 
              \hat{b}_{p+1, q}^{\dagger}\hat{b}_{p, q} + {\rm H.c.}\right )
              + \left (J_y \hat{b}_{p, q+1}^{\dagger}\hat{b}_{p, q} 
              \right. \right. \nonumber\\ 
           &+& \left. {\rm H.c.}\Big)\right] + \sum_{p, q}
              \left [\frac{U}{2}\hat{n}_{p, q}(\hat{n}_{p, q}-1) - 
               (\mu - \varepsilon_{p, q})\hat{n}_{p, q}\right],
\label{bhm}         
\end{eqnarray}
where $p$ ($q$) is the lattice site index along $x$ ($y$) direction, 
$\hat{b}_{p,q}$ ($\hat{b}^{\dagger}_{p,q}$) is the bosonic annihilation 
(creation) operator, and $\hat{n}_{p,q}$ is the number operator. 
The parameter, $J_x$ ($J_y$) is the complex hopping strength between two 
NN sites along $x$ ($y$) direction, $U$ is the on-site interaction strength. 
Here, $\mu$ is the chemical potential and $\varepsilon_{p,q}$ is the energy 
offset of the envelope potential. The envelope or confining 
potential, in the case of harmonic potential, modifies $\mu$ by the energy 
offset $\varepsilon_{p,q} = \Omega(p^2 + q^2)$, where $\Omega$ is the 
strength of the harmonic confining potential. The phase $2\pi\alpha$ in $J_x$ 
arises from the synthetic magnetic field and $0 \leqslant\alpha\leqslant 1/2$. 
It is well established that for $\alpha = 0$ the phase diagram of BHM admits 
two phases, Mott insulator (MI) and superfluid (SF) phase 
\cite{fisher_89,jaksch_98,greiner_02}. The strong on-site interaction 
limit $(J/U\ll 1)$ corresponds to the MI phase, whereas the opposite limit
$(J/U\gg 1)$ corresponds to the SF phase. The phase diagram in the 
$\mu-J$ plane consists of Mott lobes with increasing commensurate integer 
filling. And, it has been shown in previous studies that MI lobes are enlarged 
for $\alpha\neq0$ \cite{oktel_07}.


\subsection{Gutzwiller mean-field theory}

 To obtain the eigenstates of BHM, we use the mean-field approximation
\cite{sheshadri_93}. For the mean-field Hamiltonian, the annihilation 
(creation) operators in Eq.~(\ref{bhm}) are decomposed as
\begin{subequations}
\begin{eqnarray}
  \hat{b}_{p, q} &=& \phi_{p,q} + \delta \hat{b}_{p, q},\\
  \hat{b}^{\dagger}_{p, q} &=& \phi^{*}_{p, q} 
                            + \delta \hat{b}^{\dagger}_{p, q},
\end{eqnarray}
\end{subequations}
where $\phi_{p,q} = \langle\hat{b}_{p,q}\rangle$ is the SF order parameter, 
and  $\phi^{*}_{p, q} = \langle\hat{b}^{\dagger}_{p,q}\rangle$. 
Using these definitions in Eq.~(\ref{bhm}) and neglecting the second order
term in fluctuations like 
$\delta \hat{b}_{p+1,q}^{\dagger} \delta \hat{b}_{p,q}$, we obtain the
mean-field Hamiltonian of the BHM as
\begin{eqnarray}
\hat{H}^{\rm MF} &= &- \sum_{p,q}\biggr\{ \left[J_x {\rm e}^{i2\pi\alpha q}
                  \left(\hat{b}_{p+1, q}^{\dagger}\phi_{p,q} 
                  + \phi^{*}_{p + 1, q}\hat{b}_{p, q}
                  \right. \right. \nonumber \\ 
                  &-& \left. \left. \phi^{*}_{p + 1, q}\phi_{p, q}\right) 
                  + {\rm H.c.}\right] 
                  + \left[ J_y\left(\hat{b}_{p, q+1}^{\dagger}\phi_{p,q}
                  + \phi^{*}_{p, q+1}\hat{b}_{p, q}  
                  \right. \right. \nonumber \\
                  &-&\left. \left. \phi^{*}_{p, q+1}\phi_{p, q}\right) 
                  + {\rm H.c.}\right] \biggr\}
                  \nonumber \\
                  &+& \sum_{p,q} \left[\frac{U}{2}\hat{n}_{p, q}
                  (\hat{n}_{p, q}-1) 
                  - (\mu -\varepsilon_{p,q})\hat{n}_{p, q}\right]. 
\label{mf_hamil}
\end{eqnarray}
The order parameter $\phi_{p,q}$ is zero for the MI phase and finite for
the SF phase. The Hamiltonian in Eq.~(\ref{mf_hamil}) can be considered as 
the sum of the single-site Hamiltonian
\begin{eqnarray}
\hat{h}_{p,q} &=& -\left [J_x {\rm e}^{i2\pi\alpha q} 
                    \left(\phi_{p+1, q}^*\hat{b}_{p, q} 
                  - \phi^{*}_{p+1,q}\phi_{p,q} \right) + {\rm H.c.}\right ]
                          \nonumber\\ 
              &&  - \left [ J_y \left(\phi_{p, q+1}^{*} \hat{b}_{p, q} 
                  - \phi^{*}_{p,q+1}\phi_{p,q}\right) + {\rm H.c.} \right] 
                          \nonumber\\
              &&  + \frac{U}{2}\hat{n}_{p, q}(\hat{n}_{p, q}-1) - 
                 (\mu - \varepsilon_{p, q})\hat{n}_{p, q}.
   \label{ham_ss}
\end{eqnarray}
We can, therefore, diagonalize the Hamiltonian for each site separately. To 
compute the ground state of the system, we use the site dependent Gutzwiller 
ansatz. That is, the ground state of the system is the direct product of the 
ground states of all the sites,
\begin{eqnarray}
 |\Psi_{\rm GW}\rangle = \prod_{p, q}|\psi\rangle_{p, q}
                       = \prod_{p, q} \sum_{n = 0}^{N_{\rm b}}c^{(p,q)}_n
                         |n\rangle_{p, q},
 \label{gw_state}
 \end{eqnarray}
where $N_b$ is the highest occupation number basis state, $c^{(p,q)}_n$ 
are the complex coefficients of the ground state $|\psi\rangle_{p, q}$ at the
site $(p, q)$ with the normalization condition $\sum_{n} |c^{(p,q)}_n|^2$ = 1. 
Then, the SF order parameter at the lattice site ($p,q$) is
\begin{equation}
\phi_{p, q} = \langle\Psi_{\rm GW}|\hat{b}_{p, q}|\Psi_{\rm GW}\rangle 
            = \sum_{n = 0}^{N_{\rm b}}\sqrt{n} 
              {c^{(p,q)}_{n-1}}^{*}c^{(p,q)}_{n}.
\label{gw_phi}              
\end{equation}
Based on the definition of $|\Psi_{\rm GW}\rangle$ in Eq. (\ref{gw_state}), 
the MI state with density or occupancy $\rho = m$ is
\begin{equation}
  |\Psi_{\rm GW}\rangle_{\rm MI}^{m} = \prod_{p, q} 
                                 c^{(p,q)}_m |m\rangle_{p, q},
  \label{mi_state}
\end{equation}
with the condition $|c^{(p,q)}_m|^2$ = 1. Considering the above expression,
it is evident that $\phi_{p, q}$ is zero in the MI phase of the system.  
But $\phi_{p,q}$ is finite for the SF phase as more than one occupation number 
state contribute to $|\psi\rangle_{p,q}$. As the inter-site coupling is 
through $\phi_{p,q}$, and it cannot describe strongly correlated FQH states. 
For this reason, previous works have relied on CGMF~\cite{natu_16} and 
RCMF~\cite{hugel_17} to obtain FQH states in BHM. In the present work, to 
obtain ground state, the mean-field Hamiltonian is diagonalized for each 
lattice site with $N_{\rm b} = 10$ using initial guess of $\phi_{p,q}$. After 
diagonalization, the ground state is retained as the state $\ket{\psi}_{p,q}$ 
of the site in $|\Psi_{\rm GW}\rangle$. In addition, using 
$|\psi\rangle_{p,q}$ a new $\phi_{p,q}$ is computed and this cycle is 
continued till convergence. 
\begin{figure}[t]
  \includegraphics{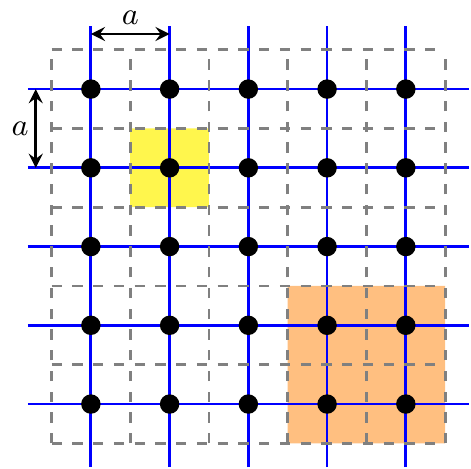}
  \caption{The solid blue lines between the lattice sites represent the
           inter-site bonds. The gray dashed lines demarcate cell around each
           lattice sites, which is used in representing cluster or 
           attributing properties to each of the lattice sites. For 
           illustration, one of the cell is highlighted in yellow and as 
           an example a 2$\times$2 cluster is identified with orange color.}
  \label{latt_cell}
\end{figure}
\begin{figure}
  \includegraphics{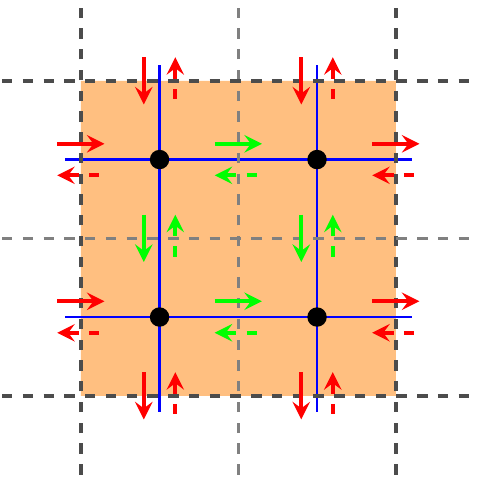}
  \caption{A $2\times 2$ cluster within the lattice. The light and bold dashed 
           lines marked boundaries of cells and cluster, respectively. The 
           solid (dashed) green colored (light gray) arrows represent the 
           exact hopping term (Hermitian conjugate) within the cluster. 
           Similarly, the solid (dashed) red colored (gray) arrows represent 
           approximate hopping term (Hermitian conjugate) across clusters with 
           one order of $\phi$ and operator. }
  \label{two_cell}
\end{figure}


\subsection{Cluster Gutzwiller mean-field theory}

From the expression of $\hat{H}^{\rm MF}$ in Eq.~(\ref{mf_hamil}), and as 
mentioned earlier it is evident that the nearest-neighbour hopping or the 
inter-site coupling is incorporated through the order parameter $\phi_{p,q}$. 
Thus, the SGMF theory does not describe the inter-site correlation very 
accurately. The CGMF remedy this by including the hopping term exactly within 
the lattice sites of a cluster. For this consider the system size is 
$K\times L$ and it is divided into $W$ clusters of size $M \times N$, that is 
$W = (K\times L)/(M\times N)$. Here, $K$, $L$, $M$, $N$, $W$ $\in \mathbb{N}$. 
A schematic description of a cluster or a representation of a cell around a 
lattice site used while representing $\rho$ are shown in Fig.~\ref{latt_cell}. 
Then, for the homogeneous systems, the limit of infinite extent is obtained 
through the periodic boundary conditions. Like in the SGMF theory, we can 
define a cluster Hamiltonian and total Hamiltonian is the sum of all the 
cluster Hamiltonians~\cite{luhmann_13}. To derive the Hamiltonian for CGMF, we 
decompose the hopping part of the Hamiltonian in two terms. First term is the 
exact hopping term for inter-site coupling within the cluster and the second 
term defines inter-site coupling for the sites at the boundary through 
mean field $\phi_{p,q}$. The Hamiltonian for a cluster can be 
written as
\begin{eqnarray}
 \hat{H}_C  &=& -\sum_{p, q \in C}'\left[\left({\rm e}^{i2\pi\alpha q}J_x 
                 \hat{b}_{p + 1, q}^{\dagger}\hat{b}_{p, q} + {\rm H.c.}\right)
                 \right. \nonumber\\
             && + \left. \left(J_y \hat{b}_{p, q + 1}^{\dagger}\hat{b}_{p, q}
                + {\rm H.c.}\right)\right]
                \nonumber\\
              &&-\sum_{p, q\in \delta C}
                \left[\left({\rm e}^{i2\pi\alpha q}J_x 
                \phi_{p+1,q}^* \hat{b}_{p, q} + {\rm H.c.}\right)
                \right. \nonumber \\
              &&+ \left. \left(J_y \phi_{p,q+1}^* \hat{b}_{p, q}
                + {\rm H.c.}\right)\right]
                \nonumber\\
              &&+\sum_{p, q \in C}
              \left[\frac{U}{2}\hat{n}_{p, q}(\hat{n}_{p, q}-1) - 
              (\mu - \varepsilon_{p, q})\hat{n}_{p, q}\right], 
\label{cg_hamil}         
\end{eqnarray}
where the prime in the summation of the first term is to indicate that 
$(p+1,q), (p,q+1) \in C$ and  $\delta C$ represents the lattice sites at the 
boundary of the cluster. The order parameter
$\phi_{p+1,q}^* = \langle\hat{b}^\dagger_{p+1,q}\rangle$ with 
$(p+1,q)\notin C$ defines the order parameter at the boundary of the 
neighbouring cluster and is required to describe the inter-cluster hopping 
along the $x$ direction. Similarly, 
$\phi_{p,q+1}^* = \langle\hat{b}^\dagger_{p,q+1}\rangle$ with 
$(p,q+1)\notin C$.  Schematically, the clusters are conveniently represented
in terms of cells. In Fig.~\ref{two_cell} the cells of a 2$\times$2 cluster
and neighbouring clusters are highlighted. 

To obtain the ground state with CGMF, we diagonalize the cluster Hamiltonian
and the ground state of the cluster in the Fock basis is
\begin{equation}
  \ket{\Psi_c} = \sum_{n_0,n_1...,n_{m'}}
                C_{n_0,n_1..,n_{m'}}\ket{n_0,n_1...,n_{m'}},
  \label{cs_state}
\end{equation} 
where $m' = (M \times N)-1$ and $n_i$ is the index of the occupation 
number state of $i$th lattice site within the cluster, 
and $C_{n_0,n_1,\ldots,n_{m'}}$ is the amplitude of the cluster Fock 
state $|n_0,n_1,\ldots, n_{m'}\rangle$. The above definition can be written
in a more compact form using the index quantum number 
$\ell\equiv \{n_0,n_1,\ldots, n_{m'}\}$ as 
\begin{equation}
   |\Psi_c\rangle = \sum_{\ell} C_\ell\ket{\Phi_c}_\ell,
\end{equation}
where $\ket{\Phi_c}_\ell$ represents the cluster basis state 
$\ket{n_0,n_1...,n_{m'}}$. The ground state of the entire $K\times L$ lattice, 
like in SGMF, is the direct product of the cluster ground states
\begin{equation}
 \ket{\Psi^c_{\rm GW}} = \prod_k\ket{\Psi_c}_k
 \label{cgw_state}
\end{equation}
where, $k$ is the cluster index and varies from 1 to
$W=(K\times L)/(M\times N)$. The SF order parameter $\phi$ is computed similar
to Eq.(~\ref{gw_phi}) as
\begin{equation}
   \phi_{p,q} = \bra{\Psi^c_{\rm GW}}\hat{b}_{p,q}\ket{\Psi^c_{\rm GW}}.
\label{cgw_phi}              
\end{equation}
As mentioned in the previous works \cite{natu_16,kuno_17}, the 
convergence is very sensitive to the initial conditions, and to accelerate 
convergence we use the method of successive over-relaxation \cite{barrett_94}.


\subsection{Exact Diagonalization Method}

\begin{figure}[t]
  \includegraphics{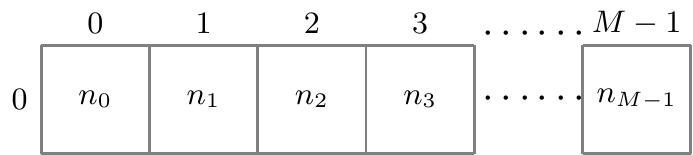}
  \caption{The $M \times 1$ row of a cluster with occupation number
           $n_0, n_1,..,n_{M-1}$. Each square box represents a lattice
           site and each of $n_i$ corresponds to $i$th lattice site in
           that row. Here, $n_i$ runs from $0$ to $N_b - 1$ for
           each lattice site.}
  \label{row}
\end{figure}

\begin{figure}[t]
  \includegraphics{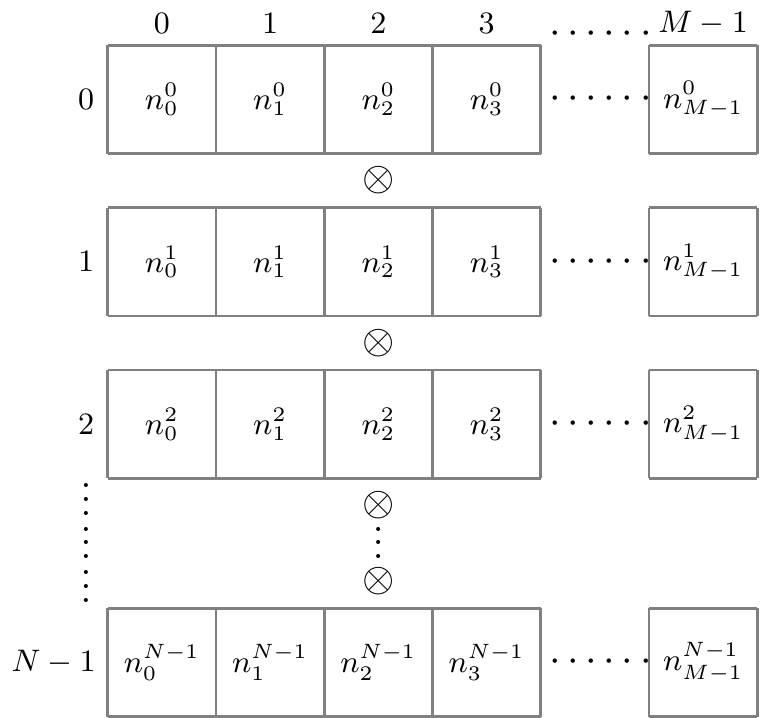}
  \caption{The $M \times N$ cluster with occupation number
           $n^j_0, n^j_1,..,n^j_{M-1}$ for $j$th row of the cluster.
           Each square box represents a lattice site and each of $n^j_i$
           corresponds to each $j$th row of cluster and $i$th lattice
           site in that row. Here, $n^j_i$ runs from $0$ to $N_b - 1$ for
           each lattice site.}
  \label{cs_row}
\end{figure}

  For an $M\times N$ lattice the computations with ED method are done with
the BH Hamiltonian 
\begin{eqnarray}
 \hat{H} &=& -\sum_{\begin{subarray}{l}
                    0\leqslant p < M \\ 0\leqslant q < N
                    \end{subarray}}\bigg[ \left( J_x{\rm e}^{i2\pi\alpha q} 
             \hat{b}_{p+1, q}^{\dagger}\hat{b}_{p, q}
             + J_y \hat{b}_{p, q+1}^{\dagger}\hat{b}_{p, q}\right) 
                          \nonumber \\
          && + {\rm H.c.}\bigg] 
             + \sum_{\begin{subarray}{l}
                     0\leqslant p < M \\ 0\leqslant q < N
                     \end{subarray}}
               \frac{U}{2}\hat{n}_{p, q}(\hat{n}_{p, q}-1).
\label{ed_bhm}         
\end{eqnarray}
Here, $\mu$ is not required as, unlike the mean field theories, the number of 
atoms is fixed and the computations are in the corresponding Hilbert space. The
Hilbert space is spanned by the states $\ket{\Psi_c}$, which like in 
CGMF can be considered as states of one $M\times N$ cluster, and the 
ground state is obtained by diagonalizing the Hamiltonian matrix. For compact
notation, we consider each $\ket{\Psi_c}$ is a direct product of $N$ row
states, and each row state is represented as
\begin{equation}
  \ket{\phi}_{m} = \prod_{i=0}^{M-1}\ket{n_i},
  \label{row_state}
\end{equation}
where, $0\leqslant i\leqslant M-1$ are lattice sites along $x$ direction,
$\ket{n_i}$ is the occupation number state at $i$th lattice site and 
$m\equiv\{n_0,n_1,...,n_{M-1}\}$ is an index quantum number of the row state. 
The schematic representation of a row state is shown in Fig.~(\ref{row}).
Thus, one of the cluster states can be written as
\begin{equation}
\ket{\Phi_c}_{\ell} =\prod_{j=0}^{N-1}\ket{\phi^j}_{m^j}
             = \prod_{j=0}^{N-1}\prod_{i=0}^{M-1} \ket{n_i^j},
\end{equation}
here, $0\leqslant j\leqslant N-1$ represent row of the cluster as shown in
Fig.~(\ref{cs_row}), and we have introduced cluster state index quantum number
$\ell \equiv \{n^0_0,n^0_1,..,n^0_{M-1}, n^1_0,n^1_1,..,n^1_{M-1},
..,n^{N-1}_0,n^{N-1}_1,..,n^{N-1}_{M-1}\}$, which is essentially equivalent 
to writing $\ell \equiv \{m^0,m^1,..,m^{N-1}\}$. In short, as shown in 
Fig.~(\ref{cs_row}) there is a hierarchy of states, the single site 
occupation number states $\ket{n^j_i}$, the row states $\ket{\phi}_m$ and 
cluster states $\ket{\Phi_c}_{\ell}$. 

Now to construct the Hilbert space, consider the total number of atoms to 
be $N_a$, and for the present work as we consider low 
density $N_a\ll M\times N$. We can, therefore, consider the occupation 
number state at each lattice site to vary from say $\ket{0}$ to $\ket{1}$, 
and consider the total number of atoms in the row states $\ket{\phi}_m$ as 
$0\leqslant \sum_i n^j_i \leqslant {\rm min} ( M, N_a)$.
However, the cluster states $\ket{\Phi_c}_{\ell}$ are direct product states of 
$\ket{\phi}_m$ such that the total number of atoms in $\ket{\Phi_c}_{\ell}$ is 
$N_a$, that is
\begin{equation}
   \sum_{i=0}^{M-1}\sum_{j=0}^{N-1}n_i^j=N_a.
\end{equation}
After diagonalizing the Hamiltonian in Eq.~(\ref{ed_bhm}) 
(for details see the appendix), we can get the ground state as
\begin{equation}
   \ket{\Psi_c} = \sum_{\ell} C_\ell\ket{\Phi_c}_\ell,
\end{equation}
where, $C_\ell$ is the coefficient of the cluster state and normalization of 
the state is ensured through the condition $\sum_\ell |C_\ell|^2 = 1$. The
normalization, however, is guaranteed as the Hamiltonian is Hermitian. As
explained in appendix, the general features of the ED method described here
can be extended to the CGMF theory to compactify the Fock space used in the 
computations.


\section{Results and discussions}

\begin{figure}[t]
  \begin{center}
  \includegraphics[width=8.0cm]{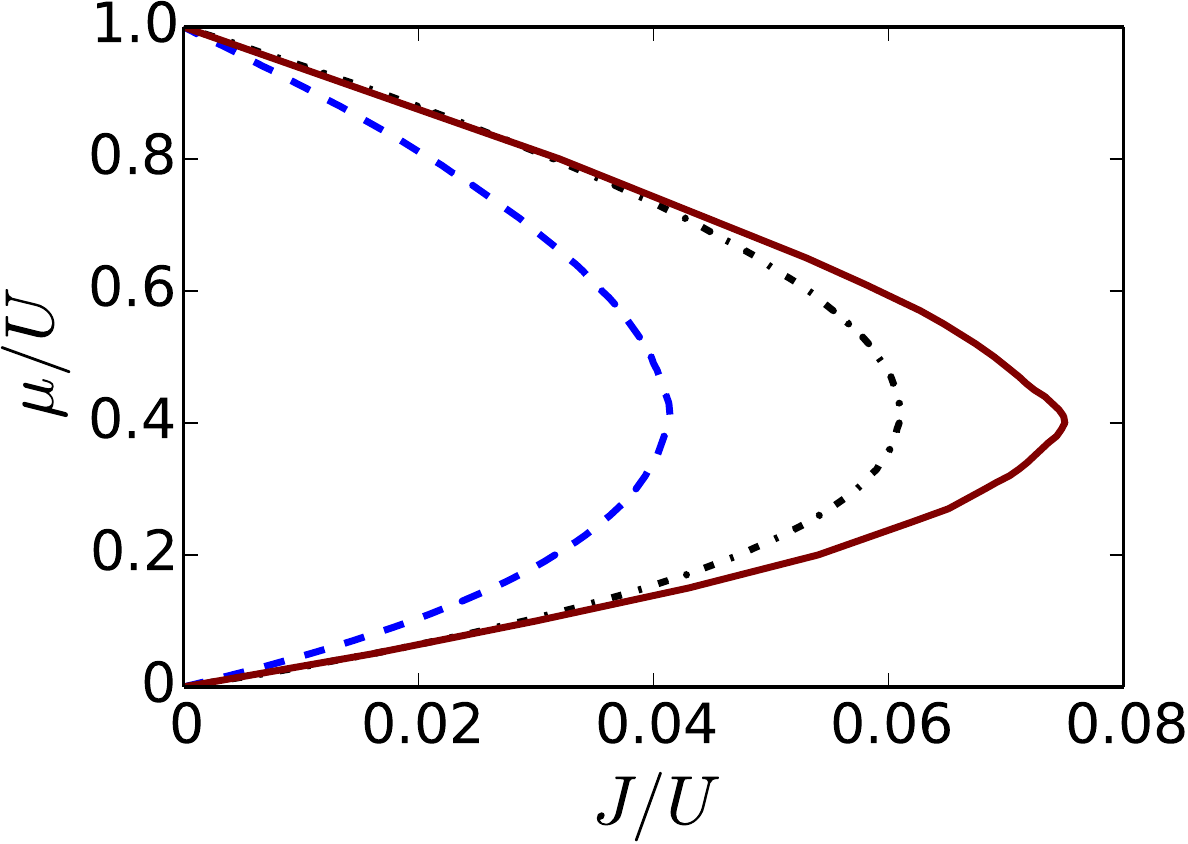}
  \caption{MI-SF phase boundary around $\rho$ = 1 Mott lobe
           with $\alpha = 0$ from SGMF theory (blue dashed line). 
           For $\alpha = 1/3$, from SGMF (black dot dashed line) and from 
           $3 \times 2$ CGMF theory (brown solid line). From the CGMF 
           calculation enhancement in the phase boundary is obtained.}
  \label{phdig}
  \end{center}
\end{figure}

  To  examine the effect of additional correlation in the CGMF compared to 
SGMF we compute the phase diagram using the two methods in presence of
artificial gauge field. For the SGMF we choose the basis size 
$N_{\rm b} = 10$, that is, the basis set of each lattice site is 
$\{|0\rangle_{p,q}$,$|1\rangle_{p,q}$, $\ldots$, $|9\rangle _{p,q}\}$. 
And, for the CGMF computations we consider a cluster basis consisting of 
single site occupation number states $\{\ket{0}, \ket{1}\}$. 
As an example, the $\rho=1$ Mott lobe obtained from SGMF and CGMF 
with $3\times 2$ clusters for $\alpha=1/3$ is shown in Fig.~\ref{phdig}. 
Based on the figure, the Mott lobe obtained from the CGMF is larger than the
SGMF. This indicates that the CGMF provides a better description of the 
strongly correlated state like the MI phase better. The other important
observation from the figure is that, the artificial gauge field enhances the 
Mott lobe. This is expected as the synthetic magnetic field induced 
cyclotron motion suppress the itinerant character of atoms in the SF phase, 
and  supports MI phase due to the localization effect \cite{niemeyer_99}. Our 
phase diagram from the SGMF theory is consistent with the results of 
Ref.~\cite{kuno_17}.

\begin{figure}
  \includegraphics[width=8.0cm]{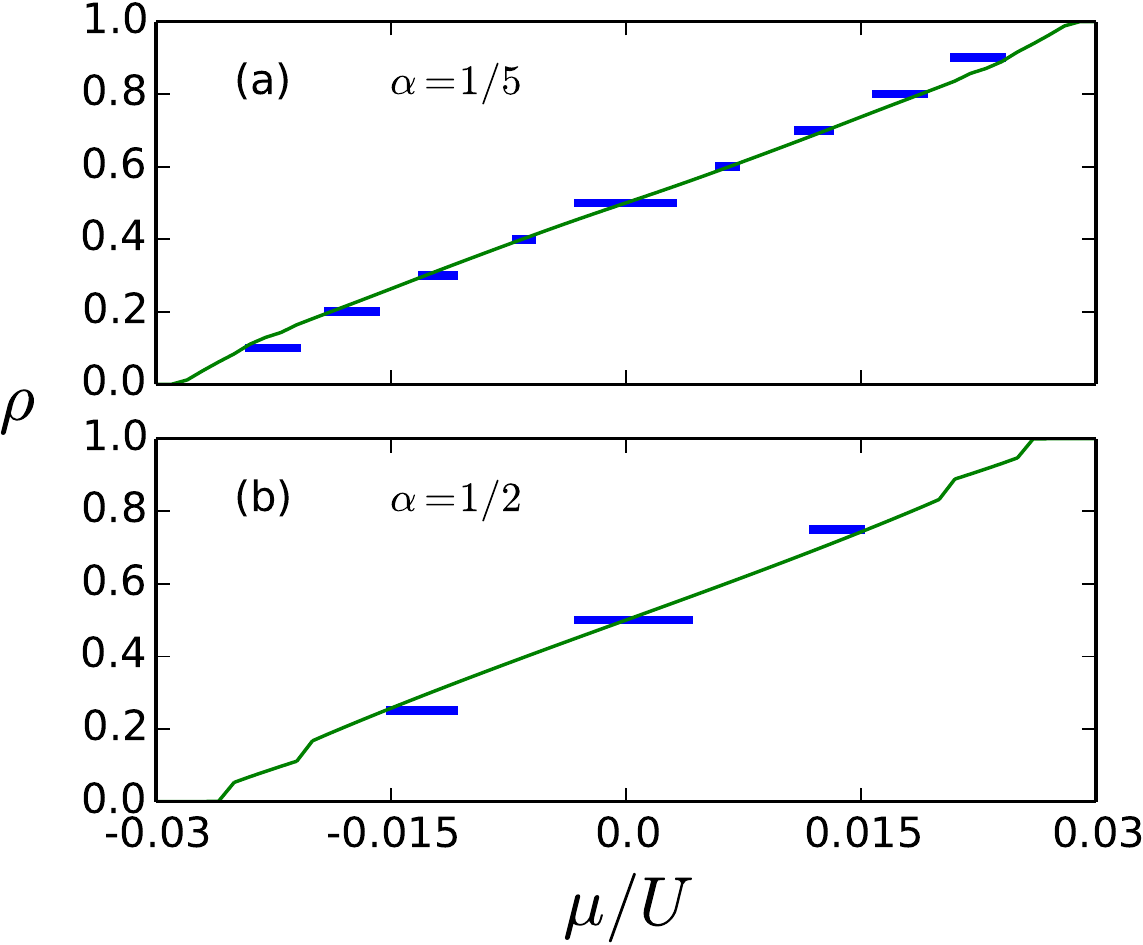}
  \caption{The number density $\rho$ with synthetic magnetic field $\alpha >0$.            The SF states are compressible, as a result, $\rho$ varies linearly 
           with $\mu$, which correspond to the green curves. The incompressible
           QH states correspond to constant $\rho$ ( blue lines)  or plateaus
           for specific values of filling factor $\nu$. (a) $\alpha = 1/5$ 
           and the plateaus correspond to $\nu = n/2, n = 1,2,..,9$.
           (b) $\alpha = 1/2$ and the plateaus correspond to 
           $\nu = 1/2$, $1$, and $3/2$}. 
  \label{kappa_alpha1b2}
\end{figure}

  The CGMF computations are done with clusters which are integer multiple of 
the magnetic unit cell. As we consider a system where the flux $\Phi$ is 
staggered along $y$-axis, for $\alpha=1/N$, a $1 \times N$ cluster forms a
magnetic unit cell. We, however, find that except for a $\pi/2$ rotation the
results are identical to $N \times 1$ cluster. This is due to the coupling of
motion along $x$ and $y$ through the interparticle interaction. The states 
obtained are classified based on the compressibilty 
$\kappa = \partial \rho /\partial \mu$, where the density 
$\rho = \sum_j\bra{\psi_c} \hat{n}_j \ket{\psi_c}/(K\times L)$.
For the QH states $\kappa=0$ or it is incompressible, and  $\kappa>0$ for 
the SF states. As a result, QH states manifest  as plateaus in $\rho(\mu)$ 
for different $\nu$ and it is linear for the SF phase. Thus, in 
Fig.~\ref{kappa_alpha1b2} the horizontal lines indicating constant
$\rho$ define the existence of QH states. Here, for simplicity and to be 
consistent with the experimental realizations we consider isotropic hopping, 
$J_x = J_y = J$, and repulsive on-site interaction, $U>0$. 


\subsection{Homogeneous system}
 Based on our results, only the QH states for $\alpha=1/4$ and $\nu=1/2$, $1$, 
$3/2$ and $2$ are ground states when $J/U\approx 0.01$, and the competing SF 
state is metastable. For the mentioned values the QH state is the ground state 
over a small range of $\mu$ centered around $-0.019U$, $-0.014U$, $-0.007U$ and
$0.000U$, respectively. For the other combination of $\alpha$ and $\nu$ the SF 
and QH states are ground and metastable states, respectively. In general, for 
different $\alpha$s, the energy difference between the SF and QH states 
$\Delta E \approx 10^{-3}$U. For the parameters of experimental interest
$U/\hbar$ = 130 Hz \cite{tai_17}  and we get  $\Delta E$ $\approx10^{-2}$nK. 
This implies stringent bounds on the thermal excitations during the state 
preparation to obtain QH states. One feature of the CGMF results which 
distinguishes the QH states from the SF states is the energy. For the QH state 
the energy decreases with increasing cluster size. For example 
the QH state of $\alpha=1/4$ with $\nu=1/2$  and $\mu=-0.02U$ has energy 
$-0.0031U$ and $-0.0046U$ with $2\times 4$, and $4\times 4$ clusters, 
respectively. Whereas for the SF state, the energy remains almost unchanged 
as it is $-0.0042U$ and $-0.0045U$, respectively. Thus, the QH state emerges as
the ground state with the $4\times 4$ cluster. Here, the key point is not the
values of the energies per se, but the importance of having better correlation
effects to obtain QH states. These trends arise from the better description 
of the hopping term with larger cluster size. Besides $\alpha=1/4$, the other
values of $\alpha$ we have studied in detail are $1/5$ and $1/2$. Results for
each of the $\alpha$ considered are described.
\begin{figure}[t]
  \begin{center}
  \includegraphics[width=8.5cm]{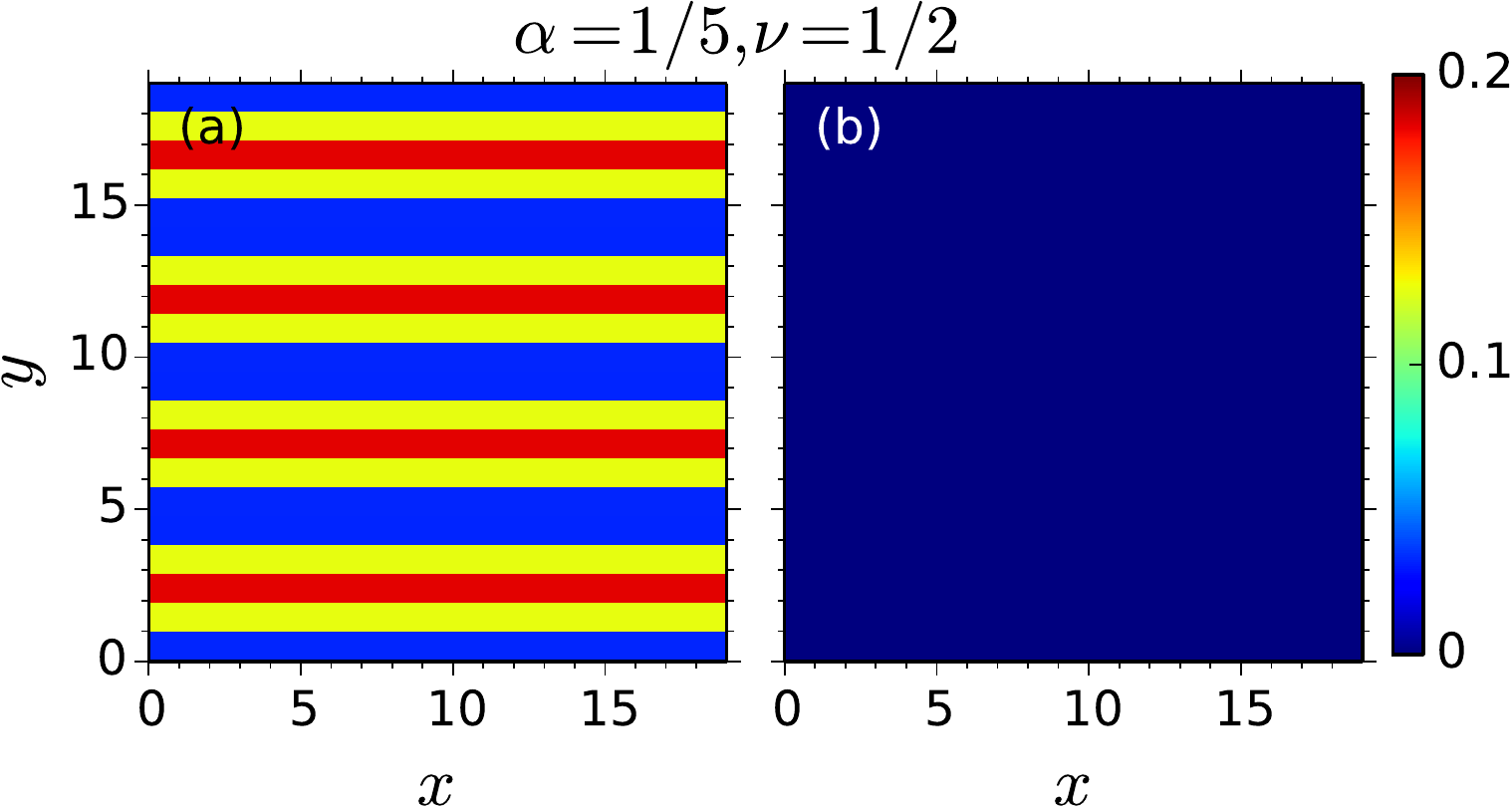}
  \caption{(a) Hall state with stripe phase for $\alpha = 1/5, \nu = 1/2$ with
      average number $\rho = 0.1$. (b) Zero SF order parameter $\phi$ for the 
      same.}
  \label{nu1b2}
  \end{center}
\end{figure}
%


\subsubsection{$\alpha=1/5$}

  For the hard-core boson limit, where $\rho < 1$, with $\alpha=1/5$, we 
obtain QH states for $\nu = n/2$, where $n=1$, $2$, $\ldots$, $9$ with 
$2\times 5$ cluster. The case of $\nu=1/2$ was reported by 
Natu {\it et al.} \cite{natu_16}, and as shown in 
Fig. \ref{nu1b2} our results are consistent. Among the 
new FQH states we have identified $\nu=3/2$, $7/2$, and $9/2$ are stripe phase 
whereas it is homogeneous for $\nu = 5/2$. In addition, we obtain stripe phase 
integer QH (IQH) states for $\nu=1$, $2$, $3$ and $4$ fillings. The other 
distinguishing feature of $\nu = 2$ and $5/2$ is that the competing SF states 
have zigzag order in $\rho$ and $\phi$. On increasing the cluster size to 
$3\times 5$ the QH states with stripe geometry are transformed to checkerboard,
and the density contrast is reduced on increasing the cluster size to 
$4\times 5$. We also obtain the same QH states but rotated by  $\pi/2$ when 
the cluster sizes are $5\times 2$, $5\times 3$ and $5\times 4$. For example 
with $5 \times 2$ cluster the stripe order is horizontal while it is 
vertical for $2 \times 5$ cluster. Considering this property of QH states, and 
noting that $1\times 5$ is the magnetic unit cell for $\alpha = 1/5$, an 
accurate description of the FQH state is possible with $5\times 5$ cluster. 
With this cluster size the operator part of the hopping term in 
Eq.~(\ref{bhm}) is exact along $x$ and $y$ axis within the cluster 
symmetrically. For example, with $2\times 5$ cluster, hopping along $x$ axis 
has contribution through mean-field after $2a$ while it is $5a$ for 
$5\times5$ cluster, where $a$ is lattice constant.


\subsubsection{$\alpha=1/4$}

  For the case of $\alpha=1/4$, we obtain QH states for $\nu = n/2$, where 
$n = 1$, $2$, $\ldots$, $7$, with $2\times 4$ and $4\times 4$ clusters. The 
FQH states for $\nu = 1/2$, $3/2$, $5/2$ have stripe order with $2 \times 4$ 
cluster, however, like in the case of $\alpha=1/5$ is transformed into 
checkerboard order with $4 \times 4$ cluster. That is, the geometry 
depends on the cluster size. Furthermore, as we increase the cluster size to 
$4 \times 8$ the FQH state with $\nu = 1/2$ filling remain 
qualitatively unchanged. For the IQH states the $\nu = 1$ and $3$ have stripe 
order with $2\times4$ cluster and checkerboard with $4\times4$ cluster. 
But, the IQH state corresponding to $\nu = 2$ has homogeneous density order. 
It must be mentioned that the thermodynamic limit, due to the coupling of 
neighbouring clusters through $\phi$, does not apply to CGMF description of 
QH states where $\phi=0$. This limits the applicability of the theory to 
finite size systems relevant to experimental realizations in optical 
lattices. On the other hand for the competing SF state a large lattice size, 
due to the finite $\phi$, corresponds to the thermodynamic limit. 
\begin{figure}[t]
  \begin{center}
  \includegraphics[width=8.5cm]{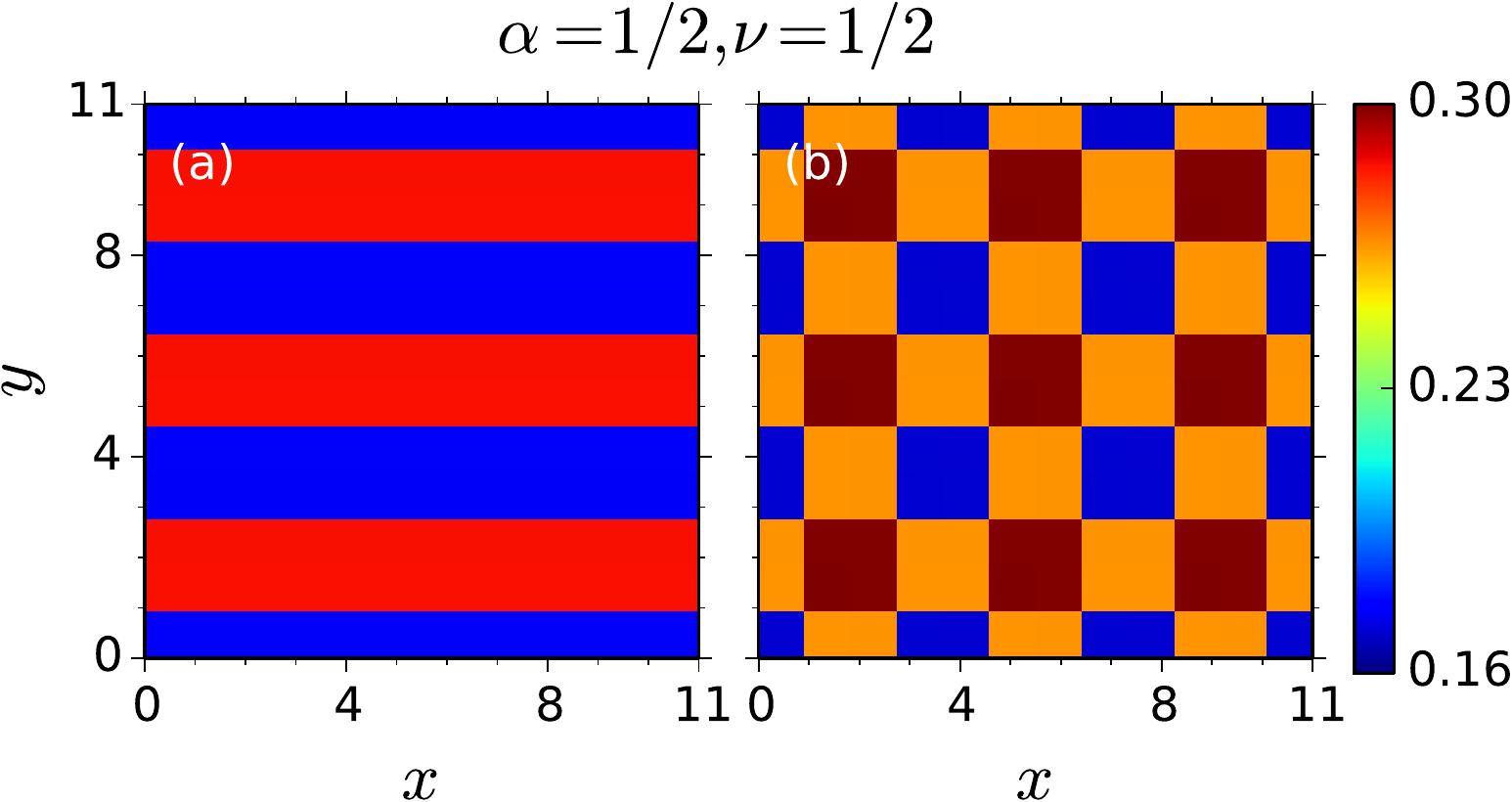}
  \caption{The variation in the lattice occupancy $\rho$ of the FQH states with
           stripe and checkerboard geometry for high flux $\alpha = 1/2$
           obtained using CGMF for the filling factor $\nu=1/2$.
           This is a metastable state, and ground state is in SF phase.
           (a) The FQH state has average number density $\rho=0.25$ with
           stripe pattern and it is obtained from $2\times 4$ cluster.
           (b) The checkerboard FQH state with the same number density
           obtained from CGMF theory with $4\times 4$ cluster.
           In both the cases the ground states, SF phase, like the FQH state
           has stripe and checkerboard geometries with  $2\times 4$ and
           $4\times 4$ cluster, respectively.}

  \label{nu1b2_1b2}
  \end{center}
\end{figure}
%


\subsubsection{$\alpha=1/2$}

For the high flux $\alpha = 1/2$, we again consider $2 \times 4$ 
and $4 \times 4$ clusters in the CGMF computations. It must be emphasized 
that $\alpha = 1/2$ is relevant to the recent experimental 
realizations \cite{aidelsburger_13, miyake_13}. For this value of $\alpha$, we 
obtain the QH states for $\nu= 1/2$, $1$, and $3/2$ from both the clusters. 
Like in $\alpha=1/5$ and $1/4$ cases, the $\nu = 1/2$ and $3/2$ FQH and 
SF states are stripe and homogeneous phases, respectively, 
with $2\times 4$ cluster. The structure of the FQH state is transformed into 
checkerboard with $4\times 4$ cluster. This transformation is
visible from the variation in $\rho$ for the case of $\nu=1/2$ as shown in 
Fig. \ref{nu1b2_1b2}. 
For $\nu = 1$ the IQH and SF states are 
homogeneous for both the cluster sizes. An important observation is, the 
homogeneous QH state is generic to $\rho=0.5$ for the values of $\alpha$ 
considered in the present work.


\subsection{Inhomogeneous system}

The simplest modification to 
the homogeneous system for comparison with experimental realizations is to 
impose hard-wall boundary conditions. This corresponds to the 2D optical 
lattice realization similar to the case of homogeneous BEC in a 
box potential \cite{gaunt_13}. With the hard-wall boundary we recover the 
QH states for all $\alpha$s described earlier, and 
energies remain unchanged. The competing SF states, on the other hand,
have higher energies with hard-wall boundary. 
In the present work the largest cluster size in the CGMF 
computations required to encapsulate one magnetic unit cell along $y$-axis and 
maintain symmetry in the exact description of hopping term is $5\times5$ for 
$\alpha=1/5$. For this reason, we focus on the properties of the QH states of
$\alpha=1/5$. The other QH states are qualitatively similar, but 
computationally less demanding. It is also to be emphasized that the results 
of single cluster with hard-wall boundary is equivalent to ED. Because with 
hard wall boundary, we do not employ the periodic boundary condition, thus 
the mean field part vanishes and Hamiltonian becomes exact.
 \begin{figure}[t]
  \begin{center}
  \includegraphics[width=8.5cm]{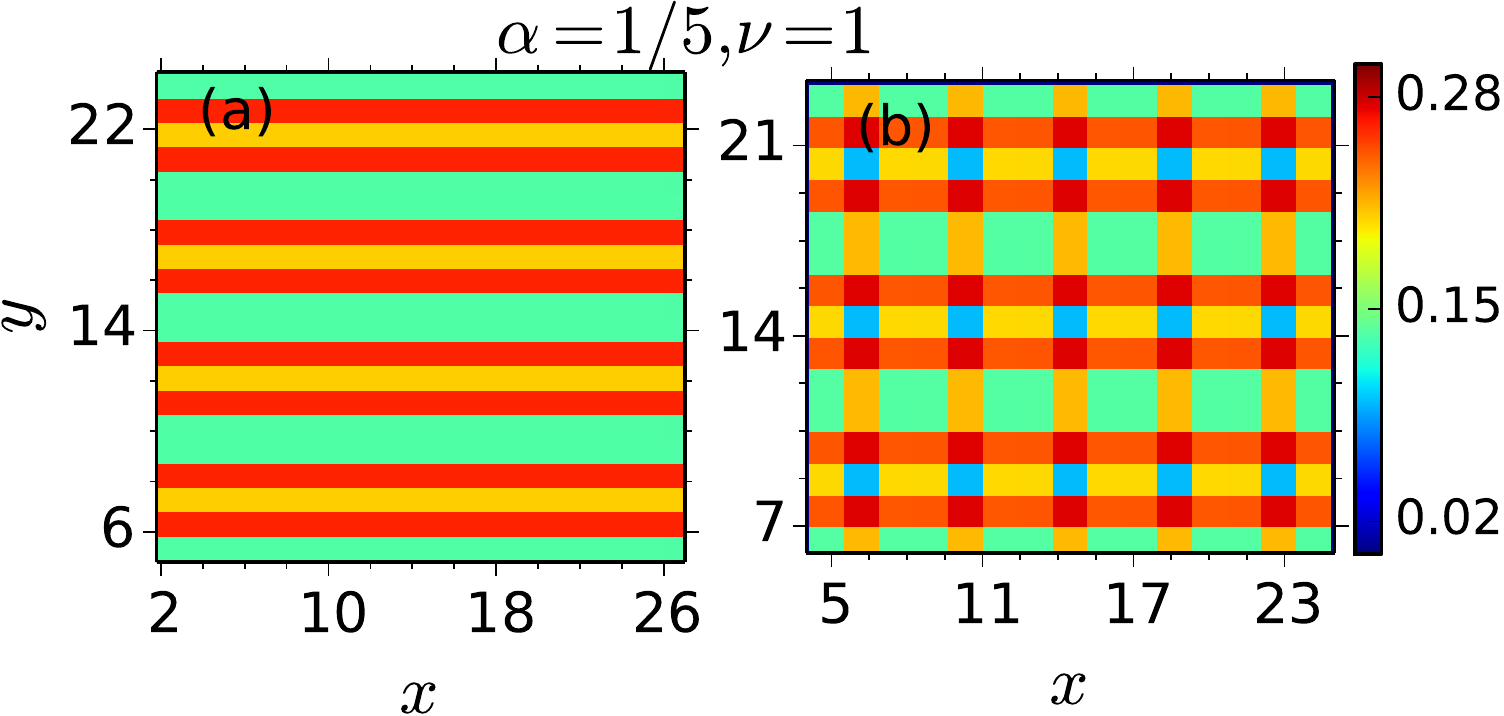}
  \caption{Density distribution of the IQH state for 
           $\alpha = 1/5$ and $\nu =1$ with hard-wall boundary. The average 
           density of atoms in this state is $\rho = 0.2$. (a) The IQH
           state has stripe geometry in the CGMF results with 
           $2\times 5$ clusters. (b) It is, however, transformed to 
           checkerboard geometry when $3\times 5$ clusters are considered 
           in the CGMF computations.}
  \label{hard_wall_nu1}
  \end{center}
\end{figure}

\begin{figure}[t]
  \begin{center}
  \includegraphics[width=8.5cm]{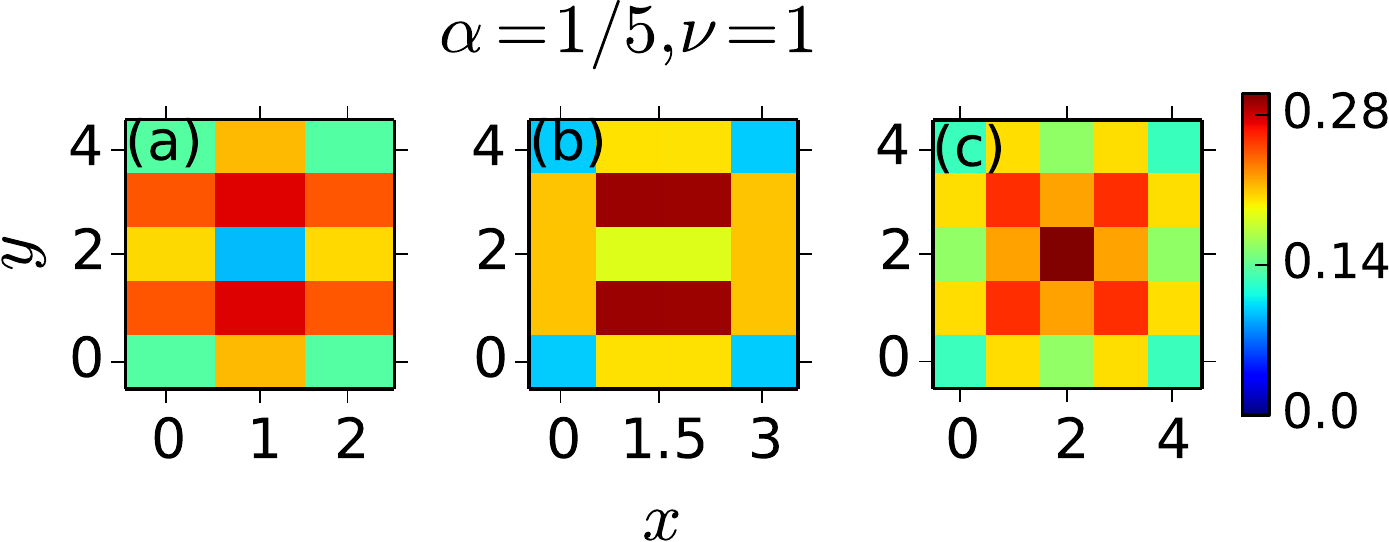}
  \caption{The variations in $\rho$ for IQH state of $\alpha=1/5$ and $\nu=1$ 
           for a single cluster of different sizes. (a) The result from 
           $3\times 5$ cluster has checkerboard pattern, and is the unit 
           cell of the large lattice shown in Fig.~\ref{hard_wall_nu1}. 
           (b) $4\times 5$ cluster has less variations in $\rho$ compared to 
           $3\times 5$. (c) $5\times 5$ cluster shows a rich variation in 
           $\rho$ and unlike in (a) and (b) the central lattice site has 
           maxima in density.
          }
  \label{qh-al1b5-nu1-cluster}
  \end{center}
\end{figure}

\begin{figure}[t]
  \begin{center}
  \includegraphics[width=7.0cm]{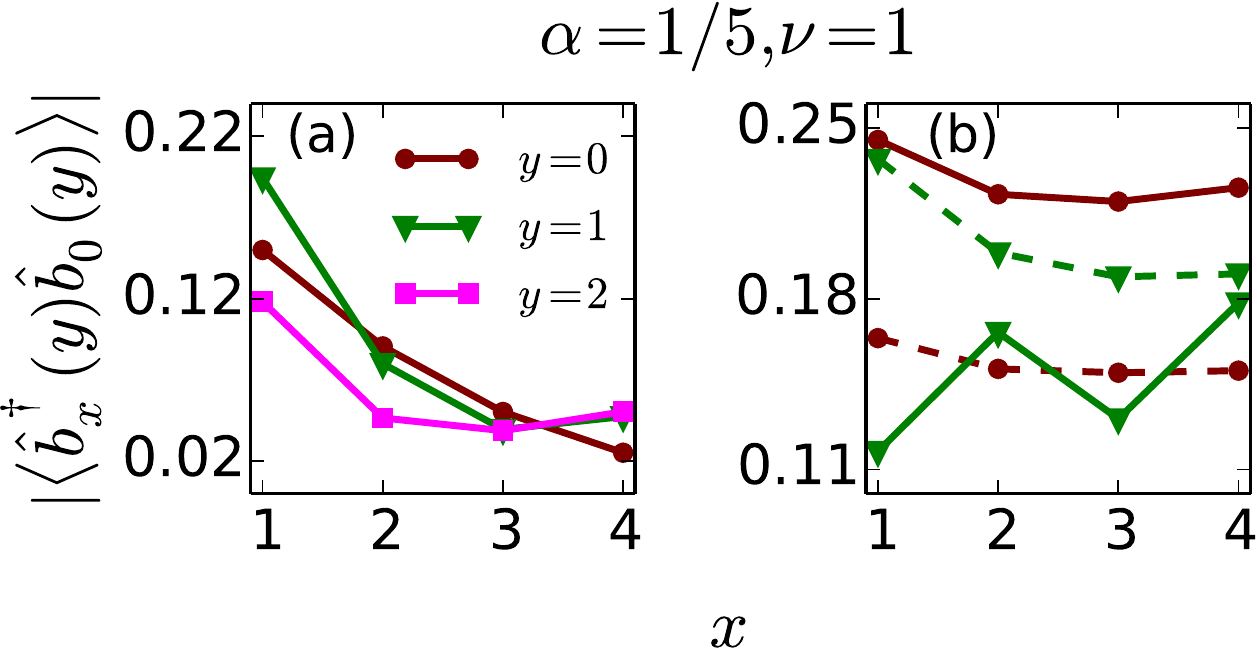}
  \caption{Two-point correlation function 
          for low flux $\alpha = 1/5$ with the  $5\times 5$ and 
          $5\times 3$ clusters for the QH and SF states, respectively. The
          correlation is calculated along the $x$ direction for the single
          cluster. Here $y = 0$ and $1$ represent the edge and bulk, 
          respectively. (a) As a characteristic feature of QH state, the 
          correlation function of the $\nu = 1$ IQH state decays 
          nonmonotonically in the bulk, and there is no difference between the 
          hard-wall and periodic boundary conditions. (b) For the corresponding
          SF state there is no trend in the bulk correlation function with 
          hard-wall boundary (solid green line with down triangle symbol), but 
          it decays monotonically at the edge (solid brown line with circle 
          symbol). With periodic boundary condition (dashed lines), the range 
          of values change, and both the bulk and edge exhibit monotonic decay 
          in correlation.
          }
  \label{corrl_func_5b3_nu_2}
  \end{center}
\end{figure}

  The IQH state for $\nu=1$ with different cluster sizes are 
shown in Fig.~\ref{hard_wall_nu1}, which has stripe geometry. Like in the 
homogeneous case, the stripe geometry is transformed into checkerboard 
geometry with $3\times 5$ cluster. However, the most important observation is 
that $\rho(x,y)$ obtained from $5\times 5$ cluster, although checkerboard in 
structure, is very different from that of $3\times 5$ and $4\times 5$, 
which are shown in Fig.~\ref{qh-al1b5-nu1-cluster}. 
An observable property to identify the QH states is the 
two-point correlation function 
$\langle\hat{b}_x^{\dagger}(y) \hat{b}_0(y)\rangle$, where the 
expectation is computed with respect to $|\psi_c\rangle$, and the results from 
the $5\times 5$ cluster are as shown in Fig.~\ref{corrl_func_5b3_nu_2}(a).
The two-point correlation function is closely related to another important
property, the one body density matrix (OBDM)~\cite{Zhang_10,Raventos_17} 
\begin{equation}
  \rho_{k,l} = \bra{\psi_c}\hat{b}_l^{\dagger} \hat{b}_k\ket{\psi_c},
  \label{obdm}
\end{equation}
where $k \equiv (x,y)$ and $l \equiv (x', y')$ are lattice indices. From
the OBDM one can compute the condensate fraction based on Penrose-Onsagar
criterion \cite{penrose_56} and von Neumann entropy 
\cite{vidal_03,kitaev_06,levin_06}. These measures are particularly relevant to 
ED method and are described while discussing the ED results.  
The correlation function, as recently proposed, could be measured with 
quantum probes \cite{Elliott_16, streif_16}.
As reported in a recent work \cite{he_17}, it can be seen from the 
figure that $\langle \hat{b}^{\dagger}_x(y) \hat{b}_0(y)\rangle$ decays as
inverse power law at the edge. However, in the bulk, as it is gaped, it 
initially shows exponential decay 
$\langle \hat{b}^{\dagger}_x(y) \hat{b}_0(y)\rangle \propto e^{-x/\xi}$
but it is power law when $x>K/2$ or on reaching the opposite edge. Here, 
$\xi$ is the correlation length of the system and as mentioned earlier, $K$ is 
the size of the cluster along $x$. For the SF state with $5\times 3$ cluster,
as seen from Fig.~\ref{corrl_func_5b3_nu_2}(b), the correlation through
the bulk does not show any nonmonotonicity. Here, we have considered 
$5\times 3$ cluster as the correlation in the bulk is not sensitive to the 
size of the cluster. 

The other envelope potential which is of experimental relevance is the harmonic
oscillator potential. Then, the energy offset
$\varepsilon_j =\Omega j^2 = \Omega (p^2 + q^2)$, $\Omega$ is the strength 
of the potential. To encapsulate the envelope
potential, we consider a larger lattice size ranging from $40\times 40$
to $80\times 80$. We, however, find that the QH 
states are absent. This is due to the nature of
$\partial \varepsilon_j/\partial j$, it monotonically increases and does not
favour incompressible phase like QH state. One possible modification is that
the beam waist $w$ of the laser beam generating the envelope potential is 
large. So that, the effective envelope potential is still a Gaussian 
$V_G = U_0 e^{-(x^2 + y^2)/w}$. Here, the amplitude of the Gaussian potential
$U_0$ is proportional to the intensity of the laser beam. With this
potential, $\partial \varepsilon_j/\partial j$ also decays exponentially and 
we find that the QH states exist for $U_0 \leqslant 10^{-3} U$. At higher 
values of $U_0 $ only the SF state is obtained from the CGMF computations.


\subsection{ED results}

With the ED computations ~\cite{Zhang_10,Raventos_17}, we 
focus our attention on the $\alpha=1/4$, which have QH states as ground state.
For this we, in particular, consider $\nu = 1/2$ FQH state with cluster sizes 
$4 \times 4$, $4 \times 8$,  and $4 \times 12$. Here, as alluded
earlier, we distinguish the QH states and SF states based on the 
Penrose-Onsager criterion \cite{penrose_56} and von Neumann entropy 
\cite{vidal_03,kitaev_06,levin_06}. For this, we compute OBDM in 
Eq.~(\ref{obdm}),
and then digonalize it. Following the Penrose-Onsager criterion, the state is 
SF if $p_m= \lambda_m^{\rm OBDM}/N\approx 1$, where $\lambda_m^{\rm OBDM}$ is 
the largest eigenvalue of the OBDM, and $N$ is the total number of atoms. 
In contrast, for the QH states $p_m < 1$. Our results are in agreement with 
this, for example, with $4 \times 4$ cluster, the values of $p_m$ are 
$0.56$ and $0.89$ for the FQH and SF states, respectively. Once the OBDM is 
diagonalized, the von Neumann entropy is defined as
\begin{equation}
  S = - \sum_i^{M}p_i \ln (p_i), 
\end{equation}
where $p_i= \lambda_i^{\rm OBDM}/N$ and $M$ is dimension of the OBDM. As the 
von Neumann entropy is a measure of entanglement, it is higher for the more 
correlated states like QH states compared to the SF states. For the states 
considered the values of $S$ are $1.0$ and $0.53$ for the FQH and SF states, 
respectively. These values indicate that the FQH state, as expected, is more 
entangled than the SF state. When the cluster size is increased 
to $4 \times 8$ the value of $p_m$ is modified to $0.26$ and $0.80$ for the 
FQH and SF states, respectively. And, the corresponding values of $S$ are 
$1.84$ and $0.95$, respectively. We also obtain similar results for the other 
QH and SF states, for example, $p_m$ is $0.33$ and $0.75$ for the QH and SF 
states respectively with $5 \times 5$ cluster for $\alpha = 1/5, \nu = 1$. 
The corresponding value of $S$ is $1.89$  and $1.20$ respectively. It is to be 
mentioned here that the QH and SF states obtained from the ED method have the 
same features, $\rho$ and $\phi$, as in CGMF results.


\section{Conclusions}
Based on the results of our studies with CGMF and ED, the 
$\alpha=1/4$ with $\nu=1/2$, $1$, $3/2$ and $2$ are the QH states which occur 
as ground state of the BHM with synthetic magnetic field, and these states
exist within a narrow range of $\mu$. For other combinations of $\alpha$ and 
$\nu$, the SF state is the ground state and the QH state exist as a metastable 
state. The experimental observation of a pure QH state needs tight control on 
the thermal excitations as the two competing states, QH and SF states, are 
nearly degenerate. The separation is only $\approx 10^{-2}$nK.  Furthermore, 
the QH state is sensitive to the nature of the envelope potential of the 
optical lattice. The QH states exist for very shallow Gaussian envelope 
potentials but cease to exist when the envelope potential is harmonic. The 
case of a box potential is the most promising experimentally realizable 
envelope potential to observe a pure QH state of BHM with synthetic magnetic 
field.

\begin{acknowledgments}

The results presented in the paper are based on the computations
using Vikram-100, the 100TFLOP HPC Cluster at Physical Research Laboratory, 
Ahmedabad, India. We thank Arko Roy, S. Gautam and S. A. Silotri for valuable 
discussions.

\end{acknowledgments}


\section*{Appendix}

\begin{figure}
  \includegraphics{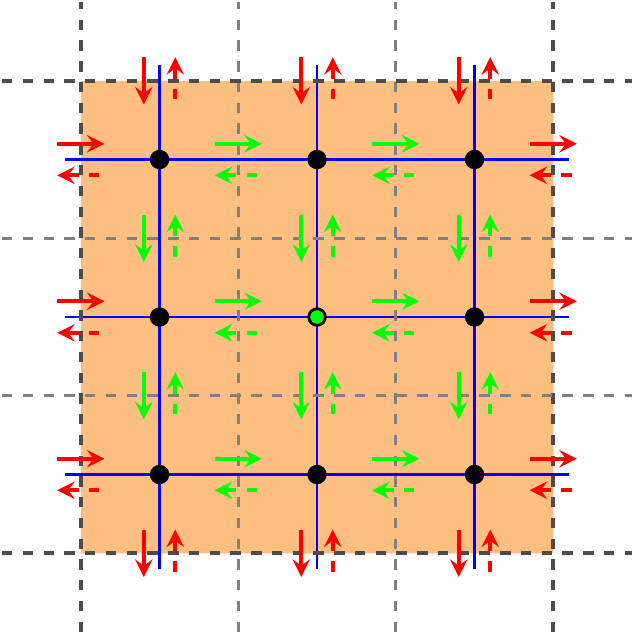}
  \caption{A $3\times 3$ cluster and form of the hopping terms between the
           lattice sites. For clarity each lattice site is represented in terms
           of cells. The light and bold dashed lines marked boundaries of
           cells and cluster, respectively. The solid (dashed) light gray 
           colored arrows represent the exact hopping 
           term (Hermitian conjugate) within the cluster. Similarly, the 
           solid (dashed) gray colored arrows
           represent approximate hopping term (Hermitian conjugate) across
           clusters with one order of $\phi$ and operator. The hopping terms
           involving the central lattice site, represented in green color, are
           all exact.
           }
  \label{three_cell}
\end{figure}

To illustrate the form of the Hamiltonian in CGMF, consider the BHM 
Hamiltonian for a $2\times 2$ cluster located at the bottom right of the 
lattice in Fig. \ref{latt_cell} is 
\begin{equation}
  \hat{h}_c = \hat{h}_{00} + \hat{h}_{10} + \hat{h}_{01}+\hat{h}_{11},
  \nonumber
\end{equation}
where $\hat{h}_{pq}$ is the single-site Hamiltonian at the $(p,q)$ lattice 
sites within the cluster. In general, if the lattice considered is 
$K\times L$, then the lattice sites are labeled along $x$ ($y$) axis as $0$, 
$1$, $\ldots$, and $K-1$ ( $0$, $1$, $\ldots$, and $L-1$). The expression of 
the single-site Hamiltonians are
\begin{eqnarray}
  \hat{h}_{00} &=& -\left ( J_x \hat{b}_{1,0}^{\dagger}\hat{b}_{0,0} 
                   + {\rm H.c} \right ) -\left(J_y \hat{b}_{0,1}^{\dagger}
                     \hat{b}_{0,0} + {\rm H.c} \right) 
                                \nonumber \\
               &&  -\left[J_x \left(\hat{b}_{0,0}^{\dagger} \phi_{K-1,0} 
                   - \phi_{0,0}^{*}\phi_{K-1,0} \right) + {\rm H.c} \right]
                                \nonumber \\
               &&  -\left[J_y \left(\hat{b}_{0,0}^{\dagger} \phi_{0,L-1} 
                   - \phi_{0,0}^{*}\phi_{0,L-1} \right) + {\rm H.c} \right]
                                \nonumber \\
               &&  +\frac{U}{2}\hat{n}_{0, 0}(\hat{n}_{0, 0}-1) - 
                    \tilde{\mu}\hat{n}_{0, 0},
                                 \\
  \hat{h}_{10} &=& -\left(J_y \hat{b}_{1,1}^{\dagger}\hat{b}_{1,0} 
                   + {\rm H.c} \right) 
                                \nonumber \\
               &&  -\left[J_x \left(\phi_{2,0}^{*} \hat{b}_{1,0} 
                   - \phi_{2,0}^{*}\phi_{1,0} \right) + {\rm H.c} \right]
                                \nonumber \\
               &&  -\left[J_y \left(\hat{b}_{1,0}^{\dagger} \phi_{1,L-1} 
                   - \phi_{1,0}^{*}\phi_{1,L-1} \right) + {\rm H.c} \right]
                                \nonumber \\
               &&  +\frac{U}{2}\hat{n}_{1, 0}(\hat{n}_{1, 0}-1) - 
                    \tilde{\mu}\hat{n}_{1, 0},
\end{eqnarray} 
\begin{eqnarray} 
  \hat{h}_{01} &=& -\left(J_x \hat{b}_{1,1}^{\dagger}\hat{b}_{0,1} 
                   + {\rm H.c} \right) 
                                \nonumber \\
               &&  -\left[J_x \left(\hat{b}_{0,1}^{\dagger} \phi_{K-1,1} 
                   - \phi_{0,1}^{*}\phi_{K-1,1} \right) + {\rm H.c} \right]
                                \nonumber \\
               &&  -\left[J_y \left(\phi_{0,2}^{*} \hat{b}_{0,1} 
                   - \phi_{0,2}^{*}\phi_{0,1} \right) + {\rm H.c} \right]
                                \nonumber \\
               &&  +\frac{U}{2}\hat{n}_{0,1}(\hat{n}_{0,1}-1) - 
                    \tilde{\mu}\hat{n}_{0,1},
                                         \\
  \hat{h}_{11} &=& -\left[J_x \left(\phi_{2,1}^{*} \hat{b}_{1,1} 
                   - \phi_{2,1}^{*}\phi_{1,1} \right) + {\rm H.c} \right]
                                \nonumber \\
               &&  -\left[J_y \left(\phi_{1,2}^{*} \hat{b}_{1,1} 
                   - \phi_{1,2}^{*}\phi_{1,1} \right) + {\rm H.c} \right]
                                \nonumber \\
               &&  +\frac{U}{2}\hat{n}_{1,1}(\hat{n}_{1,1}-1) - 
                    \tilde{\mu}\hat{n}_{1,1},
\end{eqnarray} 
where the operators and $\phi$ with index $(K-1)$ and $(L-1)$ embody  the
periodic boundary conditions along $x$ and $y$ directions, respectively. An
important point is, with the $2\times 2$ cluster none of the lattice
sites have exact representation of the hopping term. The minimal cluster
size which has exact hopping terms with respect to a lattice site is 
$3\times 3$, and the schematic diagram is shown in Fig.~\ref{three_cell}.
As seen from the figure, the hopping terms involving the central lattice site
are all exact.

For illustration of ED, consider $N_a=4$ and the size of the lattice as 
$4\times 4$. Then, the number of atoms in $\ket{\phi}_m$ can range from $0$ to 
$4$, and considering that occupation number states at each lattice sites are 
either $\ket{0}$ or $\ket{1}$, the possible row states are 
$$
  \ket{0,0,0,0}, \ket{0,0,0,1}, \ldots,\ket{1,1,1,1}.  
$$
In total there are sixteen $\ket{\phi}_m$ and an example 
of $\ket{\Phi_c}_{\ell}$ defined as direct product of four $\ket{\phi}_m$s is 
$$
  \ket{\Phi_c}_{\ell}= \ket{0,0,0,0}\otimes \ket{0,1,1,0}\otimes 
                \ket{0,0,0,1}\otimes \ket{1,0,0,0}.
$$
Thus, the number of $\ket{\Phi_c}_{\ell}$ is
$$
  ^{M \times N} C_{N_a} = ^{16}C_4 = 1820,
$$ 
which is much less than the number of states $2^{16}=65536$ required for 
computation with $4\times 4$ cluster in CGMF. 

 The essence of ED is then to compute the Hamiltonian matrix elements
between the cluster states as 
\begin{equation}
  _{\ell'}\bra{\Phi'_c}\hat{H}\ket{\Phi_c}_{\ell} 
      = \prod_{k=0}^{M-1}\prod_{l=0}^{N-1} 
      \prod_{i=0}^{M-1}\prod_{j=0}^{N-1}\bra{m_k^l}\hat{H} \ket{n_i^j},
\end{equation}
and then, diagonalize the Hamiltonian matrix to obtain the eigenvalues and 
eigenvectors. Considering that the sequence of $\ket{\Phi_c}_{\ell}$ is not 
based on symmetries, but rather based on the combinatorics of 
$\ket{\phi}_m$, the row wise computation of Hamiltonian matrix is more 
efficient. In this regard, the matrix element of the hopping term along 
$x$-axis $J_x{\rm e}^{i2\pi\alpha q}\hat{b}_{p+1, q}^{\dagger}\hat{b}_{p, q}$
can be done in the following steps:
\begin{enumerate}
   \item Compare the row states $_{m'}\bra{\phi}$ and $\ket{\phi}_m$ of 
         $_{\ell'}\bra{\Phi'_c}$ and $\ket{\Phi_c}_{\ell}$, respectively. 
         Proceed to the next step if $_{\ell'}\bra{\Phi_c}$ 
         and $\ket{\Phi_c}_{\ell}$ only differ in one of the row states, 
         say the 1st row.

   \item Consider $_{m'^1}\bra{\phi ^1}$ and $\ket{\phi ^1}_{m^1}$, and 
         compare the single site occupation number states. Proceed to the next 
         step if the difference in these two row states arise from the 
         difference in the occupation number states of two neighbouring 
         lattice sites, say 3rd and 4th lattice sites.

   \item The matrix element is nonzero and value is $\sqrt{n'_2(n'_3+1)}$ 
         if $n'_2 = n_2 + 1$ and $n'_3=n_3-1$. For the example considered,
         we have nonzero matrix element for the term $p=2$ and $q=1$. 
\end{enumerate}
In a similar way, for the example considered, the matrix element of 
the Hermitian conjugate term 
$J^{*}_x{\rm e}^{-i2\pi\alpha q}\hat{b}_{p, q}^{\dagger}\hat{b}_{p+1, q}$ 
is nonzero when the first two conditions are met and the last is modified to 
$n'_2 = n_2 - 1$ and $n'_3=n_3+1$. With slight modifications, the same approach
can be applied to compute the matrix elements of the hopping term along 
$y$-axis. For this case, two neighbouring row states should be different, 
and at the level of the lattice sites, the difference should be on the same 
column. Then, to have nonzero matrix element the occupation numbers should 
satisfy conditions equivalent of the third condition in the above chain of 
steps. The computation of the interaction Hamiltonian matrix elements is 
trivial as it is diagonal and does not require comparison of states. 

\begin{figure}
  \includegraphics{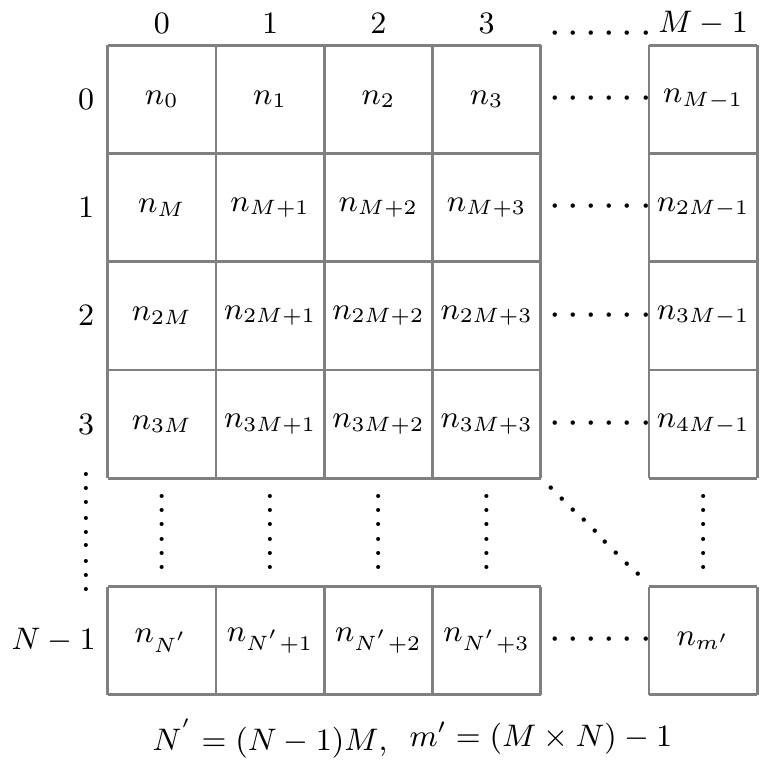}
  \caption{The $M \times N$ cluster with occupation number
           $n_0, n_1,..,n_{m'}$ at each lattice site for CGMF.
           Each square box represents a lattice site and each of $n_i$
           corresponds to each $i$ lattice site. Here, $n_i$ runs from $0$
           to $N_b - 1$ for each lattice site.}
  \label{cluster}
\end{figure}

 The general features of the hierarchical definition of states, and the
approach to compute the Hamiltonian matrix elements can also be adapted to 
the CGMF theory as well. As discussed earlier, in the CGMF theory, hopping is 
exact within the cluster but hopping at the boundary is considered via the 
mean field $\phi$. Thus, for cluster of size $M\times N$, the cluster
state defined in Eq. (\ref{cs_state}) is the direct product of the 
occupation number states at each lattice site and can be written as
\begin{equation}
  \ket{\Phi_c}_{\ell} = \prod_{i=0}^{m'}\ket{n_i},
  \label{cs_wav}
\end{equation}
where $m' = (M \times N)-1$  $i = 0$, $1$, $\ldots$, $m'$ are the lattice 
site index, with $M$ ($N$) as number of lattice sites along $x$ ($y$) 
direction, $\ell = \{n_0, n_1, \ldots, n_{m'}\}$ as defined earlier is the 
index quantum number to identify each of the cluster states uniquely. 
For illustration, the correspondence between quantum numbers and lattice 
sites is shown in Fig.~\ref{cluster}. The ground state of the CGMF 
Hamiltonian in Eq.~(\ref{cg_hamil}) is obtained by using the cluster 
state in Eq.~(\ref{cs_wav}). The Hamiltonian matrix element can 
be written as 
\begin{eqnarray}
   _{\ell'}\bra{\Phi_c}\hat{H}\ket{\Phi_c}_\ell
          &=& \prod_{j=0}^{m'}\prod_{i=0}^{m'}\bra{n'_j} 
            \hat{H}\ket{n_i}\nonumber\\
          &=& \bra{n'_0, n'_1, \ldots, n'_{m'}} \hat{H} 
            \ket{n_0, n_1, \ldots, n_{m'}}. \nonumber\\ 
\end{eqnarray}
The definition of the states and computation of the matrix elements can, 
however, be cast in terms of the row and cluster states as in ED. With this
modification, we can implement constraints on the number of atoms in the 
row and cluster states, thereby reducing the dimension of the Hamiltonian
matrix in the CGMF. The only difference from ED is, in CGMF the inter-cluster
hopping terms are linear in order parameter $\phi$ and hence, connect states
in Hilbert spaces with different total number of atoms. In other words, the
Hamiltonian matrix in CGMF is defined with respect to Fock space. Another
difference is, the diagonal terms have contribution from $\mu$. 

\bibliography{ref}{}

\end{document}